\documentclass[aps,prd,superscriptaddress,longbibliography,amsmath,amssymb,twocolumn,10pt]{revtex4-1}

\usepackage[utf8]{inputenc}
\usepackage{graphicx}
\usepackage[amssymb,Gray]{SIunits}
\usepackage{textcomp}
\usepackage{physics}
\usepackage{calc}
\usepackage{tikz}
\usepackage{comment}
\usepackage{amsmath}
\usepackage{xcolor}

\newcommand{\rmi}{\ensuremath{\mathrm{i}}}
\newcommand{\rme}{\ensuremath{\mathrm{e}}}
\newcommand{\rmd}{\ensuremath{\mathrm{d}}}
\newcommand{\bleq}{\ensuremath{\mathrel{\phantom{=}}}}
\newcommand{\nnl}{\nonumber\\}
\renewcommand{\vec}[1]{\mathrm{\mathbf{#1}}}

\newcommand{\schr}{Schr\"odinger}

\newsavebox\CBox
\newcommand\barletter[2][0.5pt]{%
  \ifmmode\sbox\CBox{$#2$}\else\sbox\CBox{#2}\fi%
  \makebox[0pt][l]{\usebox\CBox}%  
  \rule[0.5\ht\CBox-#1/2]{\wd\CBox}{#1}}

\begin{document}

%\preprint{}

\title{Causal consistency requirements for gravity-induced entanglement in near-relativistic systems with internal energy}


%\author{}
%\email[]{}
%\thanks{}
%\affiliation{}
%
\author{Linda M. van Manen}
\email[]{linda.van.manen@uni-jena.de}
\author{M.\ Kemal D\"oner}
\email[]{kemal.doner@uni-jena.de}
\author{Andr\'e Gro{\ss}ardt}
\email[]{andre.grossardt@uni-jena.de}
\affiliation{Institute for Theoretical Physics, Friedrich Schiller University Jena, Fr\"obelstieg 1, 07743 Jena, Germany}


\date{\today}

% PRL: abstract max. 600 characters
\begin{abstract}
We reconsider a thought experiment that employs the entanglement of the gravitational field with position space quantum states as a means for faster-than-light signaling. We present a protocol that includes the excitation to a higher internal energy level to increase sensitivity to gravitational phase shifts. We report that the explanations why previous versions of the thought experiment remain causally consistent are insufficient to avoid any possibility for faster-than-light signals in this case. An alternative resolution to prevent faster-than-light signaling is most reasonably the requirement for a (near) relativistic treatment. One such effect could be a decoherence channel unobserved in a nonrelativistic treatment.
\end{abstract}

%\keywords{}

\maketitle

% If in two-column mode, this environment will change to single-column
% format so that long equations can be displayed. Use sparingly.
%\begin{widetext}
%\end{widetext}

%%%%%%%%%%%%%%%%%%%%%%%%%%%%%%%%%%%%%%%%%%%%%%%%%%%%%%%%%%%%%%%%%%%%%%%
%%%  START  DOCUMENT  %%%%%%%%%%%%%%%%%%%%%%%%%%%%%%%%%%%%%%%%%%%%%%%%%
%%%%%%%%%%%%%%%%%%%%%%%%%%%%%%%%%%%%%%%%%%%%%%%%%%%%%%%%%%%%%%%%%%%%%%%

\section{Introduction}

Despite the lack of an established theory of quantum gravity, there is
a widespread belief that, as far as low-energy physics is concerned,
perturbative quantum gravity (i.e.\ the perturbative quantum field theoretical treatment of linearized Einstein gravity) provides an 
accurate model for the gravitational interactions of quantum matter.
Many questions about these interactions can be answered in direct
analogy to the electrodynamical ones. For instance, the question ``what is the gravitational field of a spatial superposition
state of a massive particle'' is immediately resolved by noticing that
the particle-field system becomes entangled, precisely as for the
electromagnetic interaction. This analogy holds for many proposed 
quantum gravitational effects, such as the recently suggested 
observation of entanglement induced by Newtonian forces~\cite{boseSpinEntanglementWitness2017,marlettoGravitationallyInducedEntanglement2017}, some gravitational decoherence effects~\cite{blencoweEffectiveFieldTheory2013}, or thought experiments about the consistency of quantum gravity~\cite{belenchiaQuantumSuperpositionMassive2018,rydvingGedankenExperimentsCompel2021}.

Classical electromagnetism and general relativity, on the other hand, are appreciably dissimilar. For instance, Maxwell's equations are linear, whereas Einstein's equations constitute nonlinear partial differential equations for the metric tensor components. A further crucial difference lies in the vector space structure of matter fields (described by sections of a vector bundle over a manifold) in contrast to the affine structure formed by the set of connections describing the gravito-inertial field~\footnote{See~\cite{giuliniCouplingQuantumMatter2022} (section 1) for a more detailed account}.
Therefore, when looking for guidance why perturbative quantum gravity so stubbornly refuses to submit to the same renormalization rules that are effective in the case of quantum electrodynamics, a promising strategy could be to consider situations featuring these exact dissimilarities between gravity and electromagnetism, both to design experimental tests and the study of internal theoretical consistency. An example is the proposition of heuristic effects of gravitational time dilation in quantum systems~\cite{zychQuantumInterferometricVisibility2011}, which results in the prediction of a genuinely gravitational decoherence effect~\cite{pikovskiUniversalDecoherenceDue2015} that cannot be explained by mere analogy to electrodynamics.

Here we discuss the potential for a novel effect of this kind: an effective two-level system, when excited while in a spatial superposition state, may exhibit gravitational decoherence at a rate determined by its virtual quadrupole distribution. This follows from a re-analysis of the causal consistency argument made by Belenchia et al.~\cite{belenchiaQuantumSuperpositionMassive2018} if applied to interference experiments with such excitable systems. As will be demonstrated, the previous arguments preventing the exploitation of gravitating quantum systems for faster-than-light signaling no longer hold in this scenario, affecting the need for an alternative resolution.

Notably, the argument applies to near-relativistic systems, in which the internal kinetic energy becomes comparable to (or supersedes) the rest mass energy. This is interesting from a fundamental level, as we predict a decoherence effect in a regime where the applicability of the usual description of decoherence within the framework of non-relativistic open quantum systems is not obvious. It may also have observational consequences, if systems can be designed in which non-gravitational decoherence effects scale at a weaker rate with the internal energy distribution, which may allow for the design of experiments in which the potential novel effect becomes the dominant decoherence channel.

We briefly review the main arguments made by Belenchia et al.~\cite{belenchiaQuantumSuperpositionMassive2018} in section \ref{sec:belenchia}, followed in section \ref{sec:internalenergy} by the analysis of the proposed modification involving internal energy levels. We discuss potential consequences in section~\ref{sec:discussion}.

\section{Causal consistency of Newtonian gravity as a which-way detector}\label{sec:belenchia}
Consider the following thought experiment~\cite{mariExperimentsTestingMacroscopic2016}: Alice and Bob are at two distant locations, separated by a distance $D$. Alice has previously prepared a superposition state $\ket{\chi} = (\ket{Q_+} + \ket{Q_-}) / \sqrt{2}$ of two mass distributions corresponding to mass quadrupoles $Q_\pm = Q_0 \pm \Delta Q$. Alice now performs an interference experiment during which the effective quadrupole is closed in time $T$. As long as no which-way information can be obtained, Alice observes an interference pattern.

Bob, on the other hand, attempts to obtain which-way information about Alice's system through a gravitational measurement by any means that are sensitive to the difference $\Delta Q$ in Alice's quadrupole moment. In the original proposal~\cite{mariExperimentsTestingMacroscopic2016, belenchiaQuantumSuperpositionMassive2018}, Bob monitors the trajectory of a freely falling test particle. The Newtonian potential at the location $D+\Delta x$ of the particle---offset by some small $\Delta x$ from Bob's location at distance $D$ from Alice---is
\begin{equation}
    \Phi_\pm = \Phi_0 - \frac{G \, Q_\pm}{(D+\Delta x)^3}
    \approx \widetilde{\Phi}_0 + 3 G \, \Delta x \frac{Q_0 \pm \Delta Q}{D^4} \,.
\end{equation}
Accordingly, after a free flight time $\tau_f$, there is a difference
\begin{equation}\label{eqn:bob-displacement}
    \delta
    = \frac{\tau_f^2}{2} \frac{\partial}{\partial x} (\Phi_+ - \Phi_-)
    = \frac{3 G \, \tau_f^2 \, \Delta Q}{D^4}
\end{equation}
between the trajectories corresponding to states $\ket{Q_\pm}$.

If Bob can resolve this displacement $\delta$, he acquires which-way information about Alice's experiment, implying that Alice will not be able to observe an interference pattern. She can then tell whether or not Bob attempted to detect the gravitational potential, which can be used as a binary communication channel between them. When $T + \tau_f < D/c$, this communication happens faster than light.

Belenchia et al.~\cite{belenchiaQuantumSuperpositionMassive2018} argue that Alice must perform her interference experiment sufficiently slowly to avoid loss of coherence through graviton emission. The energy emitted through gravitational waves is at least (see appendix~\ref{app:minimal-rad-energy})
\begin{equation}\label{eqn:grav-emission}
    E = \frac{64 G \, \Delta Q^2}{c^5 T^5} \,.
\end{equation}
Requiring $E < 2 \pi \hbar / T$, i.e. emission of not even a single graviton of wavelength $cT$ or shorter, we find
\begin{equation}\label{eqn:delta-q-condition}
   \Delta Q < \frac{\sqrt{2 \pi}}{8} \, m_P \, c^2 T^2 \,,
\end{equation} 
where $m_P$ denotes the Planck mass. Inserting this requirement into \eqref{eqn:bob-displacement} implies
\begin{equation}\label{eqn:displacement-limit}
    \delta < \frac{3 \sqrt{2 \pi}}{8} \,\frac{G \, m_P \, c^2 \, \tau_f^2 \, T^2}{D^4} < \frac{3 \sqrt{2 \pi}}{8} \, l_P \,,
\end{equation}
with $l_P$ the Planck length, and the conditions for faster-than-light signaling, $c T < D$ and $c \tau_f < D$,  were inserted in the final step.

It is generally assumed that a displacement below the Planck length---as condition \eqref{eqn:displacement-limit} would imply---cannot be detected. In a quantized theory of gravity this follows either as a consequence of spacetime vacuum fluctuations or a fundamental minimal length scale, whereas in classical general relativity it can be seen as a consequence of the fact that the localization of a quantum probe particle can only be as small as to not create a black hole singularity due to its momentum uncertainty.

\subsection{Interferometric detection}
Can Bob improve the sensitivity of his field detector over that of a free-falling test particle, such that he can gather which-way information about Alice's state in a shorter time and decohere her state in a time frame that would allow for faster-than-light signals?

An obvious candidate for a more sensitive detection would be an interferometric experiment. Assume that Bob has a spin-\textonehalf \; particle sent through a Stern--Gerlach interferometer. The initial state is the $z$-spin eigenstate $\ket{\uparrow} = (\ket{+} + \ket{-})/\sqrt{2}$ and a spatial wave function $\ket{D}$ well localized at $x=D$. The full system, including Alice's quadrupole in initial superposition $\ket{\chi}$, starts in the initial state
\begin{equation}
    \ket{\Psi(0)} = \ket{\chi} \ket{D} \ket{\uparrow} \,.
\end{equation}
Now send Bob's particle through a spin-dependent potential $F(t) x \hat{\sigma}_x$ with
\begin{equation}
    \int_0^{\tau_a} F(t) \rmd t = 0
    \,,\quad\text{and}\quad
    \int_0^{\tau_a} \int_0^t F(t') \rmd t' \rmd t = d \,,
\end{equation}
such that after an acceleration period $\tau_a$ the particle is back at its initial velocity and a position $x = D \pm d$ from the original position $D$, with sign depending on its $x$-spin component. Hence, we find
\begin{equation}
    \ket{\Psi(\tau_a)} = \frac{1}{\sqrt{2}}\ket{\chi} \left(
    \ket{D+d} \ket{+} + \ket{D-d} \ket{-}
    \right)\,.
\end{equation}
This state evolves freely for a time $\tau_f$, during which it acquires a gravitational phase (see appendix~\ref{app:grav-phase}). The phase depends on the state of Alice's quadrupole and the position of Bob's particle. After re-merging the trajectories with an inverted force profile, the wave function at the total time $\tau_t = \tau_f + 2\tau_a$ is
\begin{align}
    \ket{\Psi(\tau_t)} 
    &\sim 
      \ket{Q_+}\ket{D} \left(\ket{+} 
      + \rme^{-\rmi \gamma} \rme^{-\rmi \Gamma} \ket{-} \right)
    \nnl &\bleq
    + \ket{Q_-}\ket{D} \left(
    \rme^{-\rmi \gamma} \ket{+}
    + \rme^{-\rmi \Gamma} \ket{-} \right)
    \,,
\end{align}
up to an irrelevant global phase, and with relative phases
\begin{equation}
    \Gamma = \frac{6 G m \tau_e d}{\hbar D^4} Q_0 \,,\quad\text{and}\quad
    \gamma = \frac{6 G m \tau_e d}{\hbar D^4} \Delta Q \,,
\end{equation}
where the effective time $\tau_e$ for the optimal trajectory in appendix~\ref{app:minimal-rad-energy} is
\begin{equation}
    \tau_e = \tau_f + \frac{2}{d} \int_0^{\tau_a} x(t) \rmd t
    \approx \tau_f + 1.155 \tau_a \,.
\end{equation}
Regardless of the value of $\Gamma$, the spin parts of the wave function become orthogonal for $\gamma = \pi/2$, allowing for perfect distinguishability of Alice's states $\ket{Q_\pm}$ via a projective spin measurement by Bob.

Together with condition~\eqref{eqn:delta-q-condition} and the faster-than-light signaling condition $c T < D$ and  $c\tau_e < c \tau_t < D$, we then have
\begin{equation}\label{eqn:mdD-condition}
    m = \frac{\pi \hbar D^4}{12 G \tau_e\, \Delta Q \, d}
    > \frac{\sqrt{2 \pi} \, D}{3 \, d}\, m_P \,,
\end{equation}
which can be satisfied for a sufficiently massive particle.

\subsection{Planck scale limitations on visibility}
Let us now consider the precise evolution of the spatial wave function for Bob's particle. The wave functions for the $\ket{\pm}$ trajectories, respectively, solve the \schr\ equations
\begin{equation}
    \rmi \hbar \dot{\psi}_\pm(t,x) = -\frac{\hbar^2}{2m} \psi_\pm''(t,x) \mp F(t) x \psi_\pm(t,x) \,.
\end{equation}
The solution is given by
\begin{subequations}\begin{align}
    \psi_\pm(t,x) &= \rme^{\rmi \alpha_0(t)} \rme^{\pm \rmi \alpha(t) x}
    \psi_f(t,x \mp u(t)) \\
    \alpha_0(t) &= -\frac{m}{2\hbar} \int_0^t \dot{u}(s)^2 \rmd s \\
    \alpha(t) &= \frac{m}{\hbar} \dot{u}(t)     
\end{align}\end{subequations}
where $\psi_f$ solves the free \schr\ equation and $u(t)$ is the classical trajectory solving $m \ddot{u}(t) = F(t)$ with $u(0) = \dot{u}(0) = 0$.

Assuming a Gaussian initial state, the free solution is
\begin{equation}
    \psi_f(t,x) = \frac{\left(2 \pi \sigma^2\right)^{-1/4}}{\sqrt{1+\frac{\rmi \hbar t}{2 m \sigma^2}}} \exp(-\frac{x^2}{4 \sigma^2 \left(1+\frac{\rmi \hbar t}{2 m \sigma^2}\right)})\,.
\end{equation}
Calculating the expectation value of
\begin{equation}
    \mathcal{O} = \left(\ket{+} + \rme^{-\rmi \gamma} \rme^{-\rmi \Gamma}\ket{-}\right)
    \left(\bra{+} + \rme^{\rmi \gamma} \rme^{\rmi \Gamma}\bra{-}\right)
\end{equation}
after tracing out the spatial degrees of freedom, we find
\begin{subequations}\begin{align}
    \ev{\mathcal{O}}_{Q_+} &= 1 + A(t) \\
    \ev{\mathcal{O}}_{Q_-} &= 1 + A(t) \cos(2 \gamma)
    \intertext{with}
    A(t) &= \exp(-\frac{(u - t \dot{u})^2}{2\sigma^2}
    -2 \left(\frac{m \sigma}{\hbar} \dot{u}\right)^2) \,.
\end{align}\end{subequations}
For $\gamma = \pi/2$ and $A(\tau_t) = 1$ we find $\ev{\mathcal{O}}_{Q_-} = 0$ and, thus, the desired perfect distinguishability for Alice's quadrupole states.

In practice, we cannot perform an instantaneous measurement at $t=\tau_t$; instead, a real experiment averages over some finite interval $\Delta t$. Expanding $A(t)$ around $t = \tau_t$ to linear order for the optimal solution according to appendix~\ref{app:minimal-rad-energy} and averaging over the interval $[\tau_t-\Delta t,\tau_t]$ yields
\begin{equation}
    \ev{A} \approx 1 - \frac{15 m d^2 \, \Delta t}{2 \hbar \tau_a^2}\left(\frac{\hbar \tau_a}{2 m \sigma^2} + \frac{2 m \sigma^2}{\hbar \tau_a}\right) \,.
\end{equation}
The expression in parentheses is bounded from below with an ideal value for $\hbar \tau_a = 2 m \sigma^2$. We find the required time resolution $\Delta t$ from the condition $\ev{A} \approx 1$ as
\begin{equation}\label{eqn:delta-t-max}
    \Delta t \ll \frac{\hbar \tau_a^2}{15 m d^2}
    \lesssim 0.143 \frac{\hbar}{m c^2} \,,
\end{equation}
where in the last step we used the condition $c \,\tau_a \lesssim 1.464 \, d$ stemming from the requirement that Bob's particle moves at subluminal speed across the ideal trajectory from appendix~\ref{app:minimal-rad-energy}. Requiring that $\Delta t$ must be at least of the order of one Planck time, we have $m < 0.143 \,m_P$. Inserting this into equation~\eqref{eqn:mdD-condition} implies $d > 5.848\, D$, which means that Alice sits inside Bob's interferometer. Hence, having the Planck scale as a fundamental limitation also effectively prevents faster-than-light signaling in this case.

Note that these limitations are similar in spirit to limitations stemming from fringe visibility~\cite{rydvingGedankenExperimentsCompel2021}, although these arguments are not immediately applicable to the Stern--Gerlach interferometer discussed here.

\section{Gravitational field detection with internal energy states}\label{sec:internalenergy}
\begin{figure}
\includegraphics[scale=0.5]{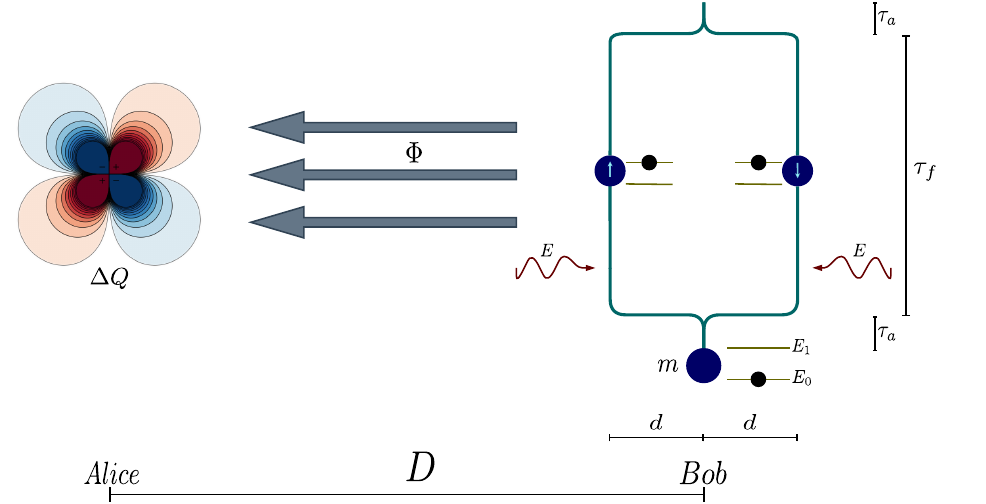}
\caption{Schematic depiction of the proposed thought experiment with interferometric detection and two internal energy levels.\label{fig1}}
\end{figure}

Thus far, we have only considered non-relativistic systems with a given, immutable rest mass $m$. Relativistically, however, the mass (inertial and gravitational) of a system is determined by its total energy $E = m c^2 + E_\text{int}$, where the internal degrees of freedom contribute in addition to the rest mass. In the more recent past, interesting effects~\cite{zychQuantumInterferometricVisibility2011,pikovskiUniversalDecoherenceDue2015} have been derived based on the idea that the contribution of internal energy elevates the mass $m$ in the \schr\ equation to an operator valued expression, resulting in correlations between internal and external degrees of freedom. Here, we do not need this operational form of the mass term, but consider the purely classical consequence that the mass parameter---through its dependence on internal energy---can acquire a time dependence.

For this purpose, assume that Bob's particle is an effective two-level system which starts in the ground state $\ket{0}$ with total energy $E_0$ and through some coherent interaction can be excited to the state $\ket{1}$ with total energy $E_1$. We discuss a concrete realization using the model of the hydrogen atom in subsection~\ref{sec:hydrogen-energy} below. For the time being, we assume that after the acceleration phase, at time $\tau_a$, Bob's system is excited from $\ket{0}$ to $\ket{1}$, and it returns to the ground state after the free flight time $\tau_f$.

Instead of equation~\eqref{eqn:mdD-condition} we then find
\begin{equation}\label{eqn:mdD-condition-int-en}
    E_1 + 1.155\,\frac{\tau_a}{\tau_f} E_0
    > \frac{\sqrt{2 \pi} \, D}{3 \, d}\, E_P \,,
\end{equation}
and condition~\eqref{eqn:delta-t-max} with the requirement that $\Delta t > t_P$ reads
\begin{equation}\label{eqn:delta-t-max-int-en}
    E_0 \ll \frac{\hbar \tau_a^2 c^2}{15 d^2 \, \Delta t}
    \lesssim 0.143 \,\frac{\hbar}{\Delta t} < 0.143 \,E_P \,.
\end{equation}
Clearly, conditions \eqref{eqn:mdD-condition-int-en} and \eqref{eqn:delta-t-max-int-en} can be simultaneously satisfied for a system where the ground state energy $E_0$ is sufficiently far below and the excited energy $E_1$ is sufficiently far above the Planck energy.

\subsection{Energy levels in a hydrogen quasiatom}\label{sec:hydrogen-energy}
Consider a generalized hydrogen atom, constituted of two massive point particles of total mass $M = m_1 + m_2$ and reduced mass $\mu = m_1 m_2 / M$ with charges $\pm q$, such that they are subject to the attractive electromagnetic Coulomb interaction and an electromagnetic potential (in Lorentz gauge and absence of further charges~\cite[ch.~11]{schiffQuantumMechanics1968}) with atom Hamiltonian
\begin{equation}\label{eqn:atom-hamiltonian}
    H_\text{atom} = \sum_{i = 1}^2 \frac{1}{2 m_i} \left( \vec p_i - q \vec A(t,\vec r_i) \right)^2 
    - \frac{k_e q^2}{\abs{\vec r_1 - \vec r_2}} \,,
\end{equation}
where $k_e = 1/(4\pi \varepsilon_0)$ is the Coulomb constant. In appendix~\ref{app:atom-light} we show the separation in center-of-mass coordinates and matrix elements due to atom-light interaction.

Introducing the Rydberg energy $E_R = k^2 q^4 \mu / (2 \hbar^2)$, we find the energy levels $E_0 = M c^2 - E_R$ for the $n=1$ ground state and $E_1 = M c^2 - E_R/4$ for the $n=2$ excited state. Assuming then $\tau_a \ll \tau_f$, such that the second term on the left-hand side of equation~\eqref{eqn:mdD-condition-int-en} becomes negligible, and taking the separation $D$ between Alice and Bob comparable to the minimum possible distance without having their systems overlap, i.e.\ $D \sim d$, we have
\begin{equation}
    \frac{3}{4\sqrt{2\pi}} (4 M c^2 - E_R) \stackrel{\text{\eqref{eqn:mdD-condition-int-en}}}{>} E_P \stackrel{\text{\eqref{eqn:delta-t-max-int-en}}}{>} 6.993 (M c^2 - E_R)
\end{equation}
which implies $E_R > 0.866\, M c^2$. From equation~\eqref{eqn:mdD-condition-int-en} we then have $M > 1.066\, m_P$. With the definition of $E_R$ we find a charge requirement of $q > 13.4\, e \, (M/\mu)^{1/4}$ (amounting to $q > 19\, e$ for $m_1 = m_2$). 

The Bohr radius achieving these bounds is
\begin{equation}
 a_0 = \frac{\hbar^2}{k \mu q^2} 
 < \frac{\hbar^2}{179.6 k e^2 \sqrt{M \mu}} 
 \lesssim \frac{\hbar^2 \sqrt{M/\mu}}{191.4 k e^2 m_P},
 \label{eqn:bohr_radius}
 \end{equation}
which can be above Planck length if the total mass $M$ is significantly higher than the reduced mass $\mu$. For the case of $m_1=m_2=m$, we find a Bohr radius of $0.356 \; l_p$, hence below the Planck mass. In the opposite limit where $m_1 \gg m_2$, we have $M \approx m_1$, and $\mu \approx m_2$. From (\ref{eqn:bohr_radius}) we get $a_0 \approx 0.71\; l_p \; \sqrt{M/\mu}$, hence to obtain $a_0 > l_p$, we need $\sqrt{M/\mu} > 1.4$  or $M > 1.95 \, \mu$.

Creating a hydrogen-like system with these parameters would certainly pose a tremendous challenge. Yet, nothing seems to prevent us from doing so \emph{in principle}.

One obvious question is whether any decay channels would disallow maintaining the excited state throughout the experiment. However, in an $n$-photon field, the rate of absorption of a single photon is $n$ times as large as the spontaneous emission rate, implying that parameters can be chosen such that the $\ket{0} \to \ket{1}$ transition happens sufficiently fast while the $\ket{1}$ state is sufficiently stable to outlast the thought experiment (cf. appendix \ref{app:photon-emi-abs}).

Addressing the question whether gravity provides a channel for spontaneous decay, noting that gravitational transitions are only allowed between hydrogen eigenstates with $\Delta l = 2$~\cite{boughnAspectsGravitonDetection2006, wanwiengEffectsGravitationalWaves2023}, we can effectively suppress any gravitational decay by considering transitions between the 1s and 2p orbitals, which are electromagnetically allowed but gravitationally forbidden. (See appendix \ref{app: graviton transition}).

\section{Discussion}\label{sec:discussion}

Within the theoretical framework we have chosen, there appears to be a parameter regime where, in principle, a two-level system that is excited only for the duration of its free flight inside a Stern-Gerlach interferometer can be used as a gravitational field sensor sensitive enough to allow for faster-than-light signaling.

Although the possibility exists that this paradoxical situation points towards a more fundamental problem with the perturbatively quantized treatment of the gravitational interaction, the more likely resolution---fostered by the experience how the original paradox has been resolved~\cite{belenchiaQuantumSuperpositionMassive2018}---is that the nonrelativistic treatment of the problem neglects effects that render the gravitational phase shift unobservable in principle. One such effect could be a decoherence channel that destroys the interferometric superposition at a rate that prevents the detection of the phase shift. Although the analysis from the perspective of a perturbatively quantized gravitational field suggests that there are no gravitational decay channels, one might expect such an effect from classical considerations:

If we had created the superposition state in Bob's Stern-Gerlach device by moving the large mass $E_1/c^2$, we would have had to account for gravitational radiation emitted during the acceleration phase. Just like for Alice's system, the emitted energy would have been determined by equation~\eqref{eqn:grav-emission} which, inserting $\Delta Q = \frac{2}{3} M d^2$ into the requirement that $E < 2 \pi \hbar/T$, yields $M < \sqrt{2\pi} 3 m_P/16 \approx 0.47 m_P$. Given that the particle is in the ground state with the much smaller mass $E_0/c^2$ during the acceleration phase, gravitational waves emitted due to the accelerated motion of the particle are negligible. Nonetheless, by depositing the de-localized photon energy at the particle's position, which is in a superposition of the trajectories at $\pm d$, a virtual mass quadrupole is created. Assuming that the creation of this virtual quadrupole results in the same sort of decoherence one would expect from the radiation of gravitational energy due to acceleration when creating a real quadrupole, the possibility for faster-than-light signaling would be efficiently suppressed. There is, however, no obvious theoretical mechanism that would explain such a decoherence effect. It is, therefore, a speculative hypothesis that we hope will trigger further theoretical analysis, as well as considerations into the possibility of experimental studies.

\appendix

\section{Trajectory with minimal gravitational radiation}\label{app:minimal-rad-energy}
We want to investigate how precisely Alice should close her quadrupole to emit the minimal amount of gravitational energy. The emission generally follows from the quadrupole formula
\begin{equation}
    \frac{\rmd E}{\rmd t} = \sum_{i,j} \frac{G}{5 c^2} \left(\frac{\rmd^3 I_{ij}}{\rmd t^3}\right)^2
\end{equation}
where
\begin{equation}
    I_{ij} = \int \rho(t,\vec r) \left(r_i r_j - \frac{r^2}{3} \delta_{ij}\right) \rmd^3 r
\end{equation}
is the mass quadrupole for the mass density distribution $\rho$ of Alice's system. We take Alice's particle to be a point particle, and assume without loss of generality that the deviation from the equilibrium position is in the $x$ direction. Since any deviation from a straight trajectory could only increase gravitational radiation, we can consider a one-dimensional trajectory $x(t)$ and the quadrupole moment
\begin{equation}
    Q(t) = I_{xx}(t) = \frac{2}{3} m x(t)^2 \,.
\end{equation}
Consequently, the total energy radiated in the form of gravitational waves is given by
\begin{equation}
    E = \frac{G}{5 c^5} \int_0^T \dddot{Q}^2 \, \rmd t
    = \frac{4 G \, \Delta Q^2}{5 c^5 T^5} S
\end{equation}
where $\Delta Q = Q(0)$ and the functional
\begin{equation}\label{eqn:s-integral}
    S = \int_0^1 \left(\xi \dddot{\xi} + 3 \dot{\xi} \ddot{\xi} \right)^2 \, \rmd \tau
\end{equation}
follows after substituting $t$ and $x$ with dimensionless equivalents and should be minimized under the constraint that $\xi(0) = 1$, $\xi(1) = 0$, and $\dot{\xi}(0) = \dot{\xi}(1) = 0$.

The Euler--Poisson equation~\footnote{For the following discussion on how to determine the minimum value for $S$, credit is greatly due to user Gon\c{c}alo's answer at \protect\url{math.stackexchange.com/q/4788286}.}
\begin{equation}
    \frac{\partial f}{\partial \xi} 
    - \frac{\rmd}{\rmd\tau} \left(\frac{\partial f}{\partial \dot\xi}\right) 
    + \frac{\rmd^2}{\rmd\tau^2} \left(\frac{\partial f}{\partial \ddot\xi}\right)
    - \frac{\rmd^3}{\rmd\tau^3} \left(\frac{\partial f}{\partial \dddot\xi}\right)
    = 0 \,,
\end{equation}
where
\begin{equation}
    f(\tau,\xi,\dot\xi,\ddot\xi,\dddot\xi) = \left(\xi \dddot{\xi} + 3 \dot{\xi} \ddot{\xi} \right)^2 \,,
\end{equation}
can be recast into the simple form
\begin{equation}
    \xi(\tau) \frac{\rmd^6}{\rmd\tau^6} \xi(\tau)^2 = 0
\end{equation}
and has the solution $\xi(\tau) = \pm \sqrt{P(\tau)}$ with 
\begin{equation}
    P(\tau) = a_0 + a_1 \tau + a_2 \tau^2 + a_3 \tau^3 + a_4 \tau^4 + a_5 \tau^5
\end{equation}
nonzero in $[0,1]$. The boundary conditions $\xi(0) = 1$, $\xi(1) = 0$, and $\dot{\xi}(0) = 0$ imply the positive solution along with $a_0 = 1$, $a_1 = 0$, and
\begin{equation}
    1 + a_2 + a_3 + a_4 + a_5 = 0 \,. \label{eqn:p-bound-cond-1}
\end{equation}
In order to meet the boundary condition $\dot\xi(1) = 0$, first note that if $\dot{P}(1) = 0$ we have
\begin{equation}
    \lim_{\tau \to 1} \dot\xi(\tau) 
    = \lim_{\tau \to 1} \frac{\dot{P}(\tau)}{2 \xi(\tau)} 
    = \lim_{\tau \to 1} \frac{\ddot{P}(\tau)}{2 \dot\xi(\tau)} \,,
\end{equation}
and must further require $\ddot{P}(1) = 0$, implying
\begin{align}
    2 a_2 + 3 a_3 + 4 a_4 + 5 a_5 &= 0 \label{eqn:p-bound-cond-2} \\
    2 a_2 + 6 a_3 + 12 a_4 + 20 a_5 &= 0 \label{eqn:p-bound-cond-3} \,.
\end{align}
The underdetermined system of equations~\eqref{eqn:p-bound-cond-1}, \eqref{eqn:p-bound-cond-2}, and \eqref{eqn:p-bound-cond-3} is solved by
\begin{equation}
    \begin{aligned}
    a_2 &= a \qquad &
    a_3 &= -10 -3a \\
    a_4 &= 15 + 3a \qquad &
    a_5 &= -6 -a
    \end{aligned}
\end{equation}
yielding $\xi_a(\tau) = \sqrt{P_a(\tau)}$ with
\begin{equation}
    \dot\xi(\tau) = \frac{(1-\tau)^2 \tau (2a-5(6+a)\tau)}{2\sqrt{(1-\tau)^3(1+3\tau+(6+a)\tau^2})}
\end{equation}
having the well-defined limit $\dot\xi(1) = 0$.

Integrating $f$ yields
\begin{equation}
    S = 180 + 60 a + 9 a^2
\end{equation}
with a minimal value of $S_\text{min} = 80$ for $a=-10/3$. The ideal trajectory is, therefore,
\begin{equation}
    x(t) = x(0) \sqrt{1 - \frac{10 t^2}{3 T^2} + \frac{5 t^4}{T^4} - \frac{8 t^5}{3 T^5}}
\end{equation}
which results in the minimal gravitational energy of
\begin{equation}
    E_\text{min} = \frac{64 G \, \Delta Q^2}{c^5 T^5} \,.
\end{equation}
The maximal magnitude velocity can numerically be obtained as $v_\text{max} \approx 1.464 x(0)/T$, which implies $x(0) < 0.683 c T$ to remain subluminal.

\section{Gravitational phases}\label{app:grav-phase}
The wave function at time $\tau_b = \tau_f + \tau_a$ is
\begin{align}
    \ket{\Psi(\tau_b)} &= \frac{1}{2} \Big(
      \rme^{\rmi\phi_{++}} \ket{Q_+}\ket{D+d}\ket{+}
    \nnl &\bleq
    + \rme^{\rmi\phi_{+-}} \ket{Q_+}\ket{D-d}\ket{-}
    \nnl &\bleq
    + \rme^{\rmi\phi_{-+}} \ket{Q_-}\ket{D+d}\ket{+}
    \nnl &\bleq
    + \rme^{\rmi\phi_{--}} \ket{Q_-}\ket{D-d}\ket{-}
    \Big)
\end{align}
with the phases ($\tau$ is the effective time, $\tau \approx \tau_f$ as long as $\tau_f \gg \tau_a$)
\begin{subequations}\begin{align}
    \phi_{++} &= \frac{m \tau}{\hbar} \Phi_+(d)
    \approx \phi_0 + \frac{\Gamma}{2} + \frac{\gamma}{2} \\
    \phi_{+-} &= \frac{m \tau}{\hbar} \Phi_+(-d)
    \approx \phi_0 - \frac{\Gamma}{2} - \frac{\gamma}{2} \\
    \phi_{-+} &= \frac{m \tau}{\hbar} \Phi_-(d)
    \approx \phi_0 + \frac{\Gamma}{2} - \frac{\gamma}{2} \\
    \phi_{--} &= \frac{m \tau}{\hbar} \Phi_-(-d)
    \approx \phi_0 - \frac{\Gamma}{2} + \frac{\gamma}{2} \,,
\end{align}\end{subequations}
where we define
\begin{equation}
    \Gamma = \frac{6 G m \tau d}{\hbar D^4} Q_0 \,,\quad\text{and}\quad
    \gamma = \frac{6 G m \tau d}{\hbar D^4} \Delta Q \,.
\end{equation}
Since the spatial states $\ket{D \pm d}$ are one-to-one correlated with the spin states $\ket{\pm}$, we can omit them for readability. Pulling out $\phi_{++}$ as a global phase, we have
\begin{align}
    \ket{\Psi(\tau_b)} 
    &= \frac{\rme^{\rmi\phi_{++}}}{2} \Big(
      \ket{Q_+} \left(\ket{+} 
      + \rme^{-\rmi \gamma} \rme^{-\rmi \Gamma} \ket{-} \right)
    \nnl &\bleq
    + \ket{Q_-} \left(
    \rme^{-\rmi \gamma} \ket{+}
    + \rme^{-\rmi \Gamma} \ket{-} \right)
    \Big) \,.
\end{align}
Reuniting the trajectories with an inverted force $F(\tau_b + t) = F(\tau_a-t)$ only affects the spatial part of the wave function, mapping $\ket{D \pm d} \to \ket{D}$.

\section{Derivation of the atom--light interaction Hamiltonian}\label{app:atom-light}
We rewrite the Hamiltonian~\eqref{eqn:atom-hamiltonian} in center-of-mass coordinates,
\begin{subequations}\begin{align}
\vec R &= \frac{1}{M} \left(m_1 \vec r_1 + m_2 \vec r_2\right) \\
\vec r &= \vec r_1 - \vec r_2 \\
\vec P &= M \dot{\vec R} = \vec p_1 + \vec p_2 \\
\vec p &= \mu \dot{\vec r} = \frac{\vec p_1 - \vec p_2}{2} - \frac{\vec P \, \Delta m}{2 M}\,,
\end{align}\end{subequations}
where $\vec P$ and $\vec p$ are the conjugate momentum for $\vec R$ and $\vec r$, respectively, and we denote the signed mass difference by $\Delta m = m_1-m_2$. Using the dipole approximation~\cite{scullyQuantumOptics1997}
\begin{equation}
    \vec A(t,\vec r_i) \approx \vec A(t,\vec R) = \vec A(t) \rme^{-\rmi \vec k \cdot \vec R}
\end{equation}
for the vector potential, writing $r = \abs{\vec r}$, and introducing the Bohr radius $a = 4 \pi \varepsilon_0 \hbar^2 / (q^2 \mu)$, the atom Hamiltonian separates
\begin{subequations}\begin{align}
    H_\text{atom} &= H_\text{cm} + H_0 + H_\text{int} \\
    H_\text{cm} &= \frac{\vec P^2}{2 M} - \frac{2 q}{M} \vec P \cdot \vec A(t,\vec R) 
    + \frac{q^2}{2 \mu} \vec A(t,\vec R)^2 \\
    H_0 &= \frac{\vec p^2}{2 \mu} - \frac{k_e q^2}{r} \\
    H_\text{int} &= -\frac{q}{\mu} \vec p \cdot \vec A(t,\vec R) \,.
\end{align}\end{subequations}
We can then neglect $H_\text{cm}$ and know the eigenstates of the hydrogen atom Hamiltonian $H_0$,
\begin{equation}
    \psi_{nlm} = \sqrt{\frac{4 (n-l-1)!}{n^4 a^3 (n+l)!}}
    \rme^{-\tfrac\rho2} \rho^l L_{n-l-1}^{2l+1}(\rho) Y_l^m(\theta,\varphi)
\end{equation}
in spherical coordinates where $\rho = 2 r / (n a)$. $L_a^b$ are the generalized Laguerre polynomials and $Y_l^m$ the spherical harmonics. Specifically the 1s, 2s, and 2p\textsubscript{0} states are
\begin{subequations}\begin{align}
    \psi_{1s}(r) &= \frac{1}{\sqrt{\pi a^3}} \rme^{-\frac{r}{a}} \\
    \psi_{2s}(r) &= \frac{2 a - r}{4 a \sqrt{2\pi a^3}} \rme^{-\frac{r}{2a}} \\
    \psi_{2p_0}(r,\theta) &= \frac{r \cos\theta}{4 a \sqrt{2\pi a^3}} \rme^{-\frac{r}{2a}} \,.
\end{align}\end{subequations}

Defining the gauge transformed wave function~\cite{scullyQuantumOptics1997}
\begin{equation}
    \phi(t,\vec r) = \rme^{-\frac{\rmi q}{\hbar} \vec A(t,\vec R) \cdot \vec r} \psi(t,\vec r) \,,
\end{equation}
neglecting terms of $\order{\vec A^2}$, and using $\dot{\vec A} = -\vec E$ in radiation gauge, we find the \schr\ equation
\begin{equation}
    \rmi \hbar \dot{\phi}(t,\vec r) = \left(H_0 - q \vec E \cdot \vec r \right) \phi(t,\vec r) \,.
\end{equation}
Hence, one obtains the familiar textbook result $$H_{int} = -q \vec{r} \cdot \vec{E} = -q\; a \rho\; E \;\hat{r} \cdot \hat{e}_z,$$ with $|r|= a \rho$. The transition matrix is then 
\begin{align}
    &\bra{\psi_{100}} H_\text{int} \ket{\psi_{21 0}} \\
    &= -q\; a E \bra{\psi_{100}}\rho \;\hat{r} \cdot \hat{e}_z \ket{\psi_{210}}.
\end{align} 
We split the hydrogenic wavefunction into the product of radial and angular states $\psi_{nlm} = R_{nl}(\rho) Y_l^m(\theta,\phi)$. The angular contribution to the matrix elements of the dipole transition is given by
\begin{align}
    \bra{l'm'} \hat{r} \cdot \hat{e}_z \ket{lm} = \bra{00} \hat{r} \cdot \hat{e}_z \ket{10} ,
\end{align}
with $$\hat{r} = \sqrt{\frac{4 \pi}{3}} (\hat{u}_1^* Y_1^1 + \hat{u}_{-1}^* Y_1^{-1}+\hat{u}_0^* Y_1^0)$$ in the spherical basis. The spherical unit vectors are defined as
$$\hat{u}_1 = - \frac{1}{\sqrt{2}}(\hat{x}+i \hat{y}) \quad \frac{1}{\sqrt{2}}(\hat{x}-i \hat{y}) \quad \hat{u}_0 = \hat{z},$$
with $\hat{u}_k^* \cdot \hat{u}_l = \delta_{kl}.$
The inner product becomes $ \hat{r} \cdot \hat{e}_z = \sqrt{\frac{4 \pi}{3}}Y_1^0$. Hence,
\begin{align}
    &\bra{00} \hat{r} \cdot \hat{e}_z \ket{10} = \sqrt{\frac{4 \pi}{3}} \bra{00} Y_1^0 \ket{10}\\
    &= \sqrt{\frac{4 \pi}{3}} \int_0^{2 \pi} \int_0^{\pi}  Y_0^{0*} Y_1^0 Y_1^0 \sin(\theta) d\theta d\phi\\
    =&  2 \pi \sqrt{\frac{4 \pi}{3}}\int_0^{\pi} \frac{1}{\sqrt{4 \pi}}\frac{3}{4 \pi} \cos^2(\theta) \sin(\theta) d\theta \\
    =& \frac{1}{\sqrt{3}},
\end{align}
with $Y_0^{0} = \frac{1}{\sqrt{4 \pi}}, \quad Y_1^0 = \sqrt{\frac{3}{4 \pi}} \cos(\theta)$.
The radial contribution is given by 
\begin{align}
    &\mathcal{R}_{1s,2p_0} \equiv \bra{1s} |\rho| \ket{2p_0}  \equiv \int_0^{\infty} R_{1s}^* \; \rho  \; R_{2p_0} \; \rho^2 d\rho\\
    =& \frac{2}{\sqrt{24}}\int_0^{\infty}    e^{-\frac{3 \rho}{2}} \rho^4 d\rho = \frac{2}{\sqrt{24}} \frac{32}{243}\int_0^{\infty} x^4 e^{-x} dx\\
    =& \frac{64}{243 \sqrt{24}}\; \Gamma(5) =  \frac{64 \sqrt{24}}{243}.
\end{align}
Then 
\begin{align}
    &\bra{\psi_{100}} H_\text{int} \ket{\psi_{21 0}} \\
    &= -q\; a E \bra{00} \hat{r} \cdot \hat{e}_z \ket{10} \;  \bra{1s} |\rho| \ket{2p_0}\\
    =& -q\; a E \; \frac{64}{243} \;   \frac{\sqrt{24}}{\sqrt{3}} = - \frac{128}{243} \sqrt{2} \; q\; a E \;.
\end{align}

\section{Photon absorption and emission}\label{app:photon-emi-abs}

We have introduced the interaction Hamiltonian of an atom and external electromagnetic radiation as $H_\text{int} = \frac{-q}{\mu} \vec p \cdot \vec A(t, \vec R)$ by considering the electromagnetic radiation as a classical field. However, this treatment is only sufficient for large numbers of photon-atom interactions, hence it can be used for absorption and stimulated emission, although it can not be used for spontaneous emission. Therefore, we consider the quantization of the electromagnetic field. The quantized field $\hat{\vec{A}}(t,\vec{R})$ is given by

\begin{equation}\label{eqn:e.e.r_qua}
    \hat{\vec A}(t, \vec R) = \sum_{\vec k} \sum_{\lambda = 1}^2 \sqrt{\frac{\hbar}{2 \epsilon_0 w_k V}} \bigg[ \hat{a}_{\lambda, \vec k} e^{i(\vec k \cdot r - wt)} \vec{\epsilon}_{\lambda} + c.c. \bigg],
\end{equation}
where $\hat{a}_{\lambda, \vec{k}}$ is the photon annihilation operator, $\omega_k$ the associated frequency, and $\epsilon_{\lambda}$ is the polarization vector. Accordingly,
\begin{align}
    H_\text{int} &= \frac{q}{\mu} \sum_{\vec{k},\lambda} \sqrt{\frac{\hbar}{2 \epsilon_0 w_k V}} \vec \epsilon_{\lambda} \cdot \vec p \big[ \hat{a}_{\lambda, \vec k} \,e^{i(\vec k \cdot \vec r - w_kt)} + c.c.\big].
\end{align}
If we consider the atom in initial and final state $\ket{\psi_i}$ and $\ket{\psi_f}$, then the transition matrices for emission and absorption are respectively
\begin{align}
    &\bra{\psi_f, n_{\lambda,\vec k}-1} \frac{q}{\mu} \sqrt{\frac{\hbar}{2 \epsilon_0 w_k V}} \vec \epsilon_{\lambda} \cdot \vec p \big[ \hat{a}_{\lambda, \vec k} \,e^{i(\vec k \cdot \vec r - w_kt)}\big] \ket{\psi_i,n_{\lambda, \vec k}}\\
     &\bra{\psi_f, n_{\lambda,\vec k}+1} \frac{q}{\mu} \sqrt{\frac{\hbar}{2 \epsilon_0 w_k V}} \vec \epsilon_{\lambda}^* \cdot \vec p \big[ \hat{a}^{\dagger}_{\lambda, \vec k} \,e^{-i(\vec k \cdot \vec r - w_kt)}\big] \ket{\psi_i,n_{\lambda, \vec k}}.
\end{align}
By using Fermi's Golden rule we find the associated transition rates 
\begin{align}\label{eqn:tra_emi_qua}
    \Gamma_{i\rightarrow f}^\text{emi} &= \frac{\pi q^2 w^2_\text{fi}}{ \epsilon_0 \mu^2 w_k V} (n_{\lambda, k} + 1) \abs{  \bra{\psi_f}\vec{ \epsilon}_{\lambda}^* \cdot  \vec p \ket{\psi_i}}^2 \nnl
    &\times \delta(E_f - E_i + \hbar w_k) \\ \label{eqn:tra_abs_qua}
    \Gamma_{i\rightarrow f}^\text{abs} &= \frac{\pi q^2 w^2_\text{fi}}{\epsilon_0 \mu^2 w_k V} (n_{\lambda, k}) \abs{  \bra{\psi_f} 
 \vec{ \epsilon}_{\lambda} \cdot\vec p \ket{\psi_i}}^2 \nnl &\times \delta(E_f - E_i - \hbar w_k),
\end{align}
where we have used $\hat{a}_{\lambda,\vec{k}} \ket{n_{\lambda,\vec{k}}} = \sqrt{n_{\lambda,\vec{k}}}  \ket{n_{\lambda,\vec{k}}-1}$, $\hat{a}^{\dagger}_{\lambda,\vec{k}} \ket{n_{\lambda,\vec{k}}} = \sqrt{n_{\lambda,\vec{k}}+1}  \ket{n_{\lambda,\vec{k}}+1}$, and the dipole approximation, i.e., $\vec{k} \cdot \vec{r} \approx 0$.
The equation (\ref{eqn:tra_emi_qua}) shows a non zero rate for $n_{\lambda,\vec k}=0$, i.e., the vacuum. This process is known as spontaneous emission which can occur without radiation present. Hence, for spontaneous emission we have a rate
\begin{equation}\label{eqn:tra_spo_qua}
    \Gamma_{i\rightarrow f}^\text{spo} = \frac{\pi w^2_\text{fi}}{\epsilon_0  w V} \abs{\vec \epsilon_{\lambda}^* \cdot \vec d_{\text{fi}}}^2 \delta(E_f - E_i - \hbar w)
\end{equation}
with $\vec d_{fi} = -q \vec r$ the electrics dipole moment, and we have set $\frac{q}{\mu} \vec A(\vec R,t) \cdot \vec p = q \vec E(\vec R,t) \cdot \vec r$. Equation (\ref{eqn:tra_spo_qua}) shows the transition rate of a photon's emission along a specific direction per unit time. For the total rate, we integrate over all possible directions and replacing the discrete frequencies with the quasi-continuum case. i.e., $d^3 n = \frac{V w^2}{(2 \pi c)^3} d\Omega dw$. This gives the total transition rate
\begin{equation}\label{eqn:tra_tot_spo}
    \Gamma_{i\rightarrow f}^\text{spo} = \frac{ w^3 q^2}{2 \pi \epsilon_0 \hbar c^3} \abs{\bra{ \psi_f} \vec r \ket{\psi_i}}^2   
\end{equation}
where $w = (E_f - E_i)/\hbar$. For an understanding of the role of the number of photons, we can immediately see that for large $n_{\lambda,k}$, the rate of absorption and stimulated emission grows linearly with $n_{\lambda,k}$, while the spontaneous emission is unaffected. This explicitly shows us that increasing $n_{\lambda,k}$ directly enhances the probability of absorption and stimulated emission and reduces the mean lifetime, $\tau$, of these transitions:
\begin{equation}
    \tau = \frac{1}{\Gamma_{i\rightarrow f}(n_{\lambda,k})}.
\end{equation}

\section{Derivation of the atom--gravity interaction Hamiltonian}
We follow the discussion by Boughn and Rothman~\cite{boughnAspectsGravitonDetection2006}, starting with the matter-gravity interaction Hamiltonian density
\begin{equation}
    \mathcal{H}_\text{int} = -\mathcal{L} = \frac{1}{2} h_{\mu\nu} T^{\mu\nu} \approx \frac{1}{2} h_{00} T^{00}
\end{equation}
in a local inertial frame, where $T^{00} = c^2 \rho$ is proportional to the mass density, given by $\rho = \sum_n m_n \delta(r-r_n)$ for a point mass system. In Fermi normal coordinates, one has
\begin{equation}
    h_{00} = - R_{0j0k} r^j r^k \,,
\end{equation}
where the Riemann tensor is a gauge invariant quantity. We can evaluate this expression to lowest order in $h^{TT}_{\mu\nu}$ in the TT gauge:
\begin{equation}
    h_{00} = \frac{1}{2 c^2} \partial_t^2 h^{TT}_{jk} r^j r^k \,.
\end{equation}
For an external gravitational field from a source far away, we obtain the vacuum solution for $h_{jk}^{TT}$, which is a gravitational wave with wave vector $\vec k$, amplitude $h$,  and the polarization tensor $e_{ij}$. The polarization tensor satisfies $e_{ij}k^j =0$, \;$e_{ij} = e_{ji}$,\; $e_{\;i}^i = 0$, and is normalized by $e_{ij}e^{ij} = 2$. We have
\begin{equation}
    h^{TT}_{jk} = -2 h e_{jk} \sin(\vec k \cdot \vec r - \omega t) \,,
\end{equation}
where $\omega = c \abs{\vec k}$, and we make use of the dipole approximation $\vec k \cdot \vec r \approx 0$. Furthermore, we rewrite the Hamiltonian in relative and center-of-mass coordinates, where we neglect the center-of-mass contribution. Thus
\begin{equation}\label{eqn:gw-interaction-ham}
    H_\text{int}(\vec k, t) = \frac{h}{2} \frac{\mu}{M} \omega^2 \sin(\omega t) e_{jk} r^j r^k \,.
\end{equation}

Quantizing the gravitational waves, following the usual procedure, the Hamiltonian takes the form~\cite{boughnAspectsGravitonDetection2006}
\begin{align}
    H_\text{int} &= \frac{\mu}{4 Mc} \sqrt{\frac{16 \pi G \hbar}{V}} \sum_{\vec k, \sigma}
    \sqrt{\omega^3} \Big[ \hat{a}_{\vec k,\sigma} \rme^{\rmi(\vec k \cdot \vec r - \omega t)}
    \nnl &\bleq
    + \hat{a}^\dagger_{\vec k,\sigma} \rme^{-\rmi(\vec k \cdot \vec r - \omega t)} \Big]
    e^{\vec k,\sigma}_{ij} x^i x^j \,,
\end{align}
where $\sigma = \pm 1$ denotes the two possible polarizations of the quantized waves. Direct comparison with equation~\eqref{eqn:gw-interaction-ham} shows that
\begin{equation}
    h = \sqrt{\frac{16 \pi G \hbar}{V \omega c^2}} \,.
\end{equation}
The matrix elements between an initial state $\ket{\psi_i}\ket{0}$ with a final state $\ket{\psi_f}\ket{1}$ including a graviton with wave vector $\vec k$ are then given by
\begin{multline}\label{eqn:matrix-element-quantized}
    \bra{1}\bra{\psi_f} H_\text{int} \ket{\psi_i}\ket{0} 
    \\ =
    \frac{\mu}{Mc} \sqrt{\frac{\pi G \hbar \omega^3}{V}} 
    \rme^{\rmi\omega t} \bra{\psi_f}
    e^{\vec k,\sigma}_{ij} x^i x^j \ket{\psi_i}.
\end{multline}

\section{Graviton transitions}\label{app: graviton transition}
Let $\alpha = (n,l,m)$ be a multi-index denoting the energy eigenstates of $H_0$. We can evaluate the transition amplitudes in first-order perturbation theory~\cite{schiffQuantumMechanics1968}:
\begin{align}
    a^{(1)}_{\alpha\beta}(t) &= -\frac{\rmi}{\hbar} \int_0^{t} \rmd t' \, 
    \bra{\psi_\alpha} H_\text{int}(\vec k,t') \ket{\psi_{\beta}}
    \rme^{\rmi \omega_{\alpha\beta} t'} \nnl
    &= -\frac{\rmi}{2 \hbar} \frac{\mu}{M} \omega^2 \sqrt{\frac{16 \pi G \hbar}{V \omega c^2}}
    \bra{\psi_\alpha} e_{jk} x^j x^k \ket{\psi_{\beta}} \nnl
    &\bleq \times \int_0^{t} \rmd t' \, \sin(\omega t')
    \rme^{\rmi \omega_{\alpha\beta} t'} \nnl
    &= \frac{\rmi \mu}{Mc} \sqrt{\frac{\pi G \omega^3}{V \hbar}} 
    \bra{\psi_\alpha} e^{\vec k,\sigma}_{jk} x^j x^k \ket{\psi_{\beta}} \nnl
    &\bleq \times 
    \left( \frac{\rme^{\rmi (\omega_{\alpha\beta}+\omega) t}-1}{\omega_{\alpha\beta}+\omega}
    - \frac{\rme^{\rmi (\omega_{\alpha\beta}-\omega) t}-1}{\omega_{\alpha\beta}-\omega} \right) \,,
\end{align}
where we made use of the dipole approximation $\vec k \cdot \vec r \approx 0$ and $\hbar\omega_{\alpha\beta} = E_\alpha - E_\beta$ is the energy difference between the two states. Recall that the energy difference in the hydrogen atom is $\hbar\omega_{nn'} = E_R (1/n^2 - 1/{n'}^2)$ with $E_R = \hbar^2 / (2 m a^2)$ the Rydberg energy.

Whenever the matrix elements $\bra{\psi_\alpha} H_\text{int} \ket{\psi_{\beta}}$ between two states vanish, a single graviton transition is not possible. However, the transition can still be made via an intermediate state. The transition probabilities are given by  the absolute value squared of the second-order term of the Dyson series (again for $t>t_0$):
\begin{widetext}
\begin{align}
    a^{(2)}_{\alpha\beta}(t) =& -\frac{1}{\hbar^2} 
    \int_0^t \rmd t' \int_0^{t'} \rmd t'' \, \sum_{\gamma}
    \bra{\psi_\alpha} H_\text{int}(\vec k_2,t') \ket{\psi_{\gamma}}
    \bra{\psi_\gamma} H_\text{int}(\vec k_1,t'') \ket{\psi_{\beta}}
    \rme^{\rmi \omega_{\alpha\gamma} t' + \rmi \omega_{\gamma\beta} t''} \nnl
    =& -\frac{1}{\hbar^2} 
    \sum_{\gamma} \int_0^t \rmd t' \, 
    \bra{\psi_\alpha} H_\text{int}(\vec k_2,t') \ket{\psi_{\gamma}}
    \rme^{\rmi \omega_{\alpha\gamma} t'}
    \int_0^{t'} \rmd t'' \,
    \bra{\psi_\gamma} H_\text{int}(\vec k_1,t'') \ket{\psi_{\beta}}
    \rme^{\rmi \omega_{\gamma\beta} t''} \nnl
    =& - \frac{\mu \omega_1^2}{2M \hbar^2} \sqrt{\frac{16 \pi G \hbar}{V \omega_1 c^2}}
    \sum_{\gamma} \bra{\psi_\gamma} e^{\vec k_1,\sigma_1}_{ij} x^i x^j \ket{\psi_{\beta}}
    \int_0^t \rmd t' 
    \bra{\psi_\alpha} H_\text{int}(\vec k_2,t') \ket{\psi_{\gamma}}
    \rme^{\rmi \omega_{\alpha\gamma} t'} \hspace{-5pt}
    \sum_{\tilde\omega_1 \in \pm \omega_1} \hspace{-5pt}
    \mathrm{sign}(\tilde\omega_1) \frac{\rme^{\rmi (\omega_{\gamma\beta}+\tilde\omega_1) t'}-1}{\omega_{\gamma\beta}+\tilde\omega_1}
    \nnl
    =& -\frac{4 \mu^2 \pi G}{\hbar VM^2 c^2}
    \sqrt{\omega_1^3 \omega_2^3} 
    \sum_{\gamma} 
    \bra{\psi_\alpha} e^{\vec k_2,\sigma_2}_{kl} x^k x^l \ket{\psi_{\gamma}}
    \bra{\psi_\gamma} e^{\vec k_1,\sigma_1}_{ij} x^i x^j \ket{\psi_{\beta}}
    \nnl &\bleq \times
    \int_0^t \rmd t' \, 
    \sin(\omega_2 t')  
    \rme^{\rmi \omega_{\alpha\gamma} t'}
    \left( \frac{\rme^{\rmi (\omega_{\gamma\beta}+\omega_1) t'}-1}{\omega_{\gamma\beta}+\omega_1}
    - \frac{\rme^{\rmi (\omega_{\gamma\beta}-\omega_1) t'}-1}{\omega_{\gamma\beta}-\omega_1} \right) \nnl
    =&  \frac{2 \mu^2 \pi G}{ \hbar VM^2 c^2} 
    \sum_{\gamma} 
    \bra{\psi_\alpha} e^{\vec k_2,\sigma_2}_{kl} x^k x^l \ket{\psi_{\gamma}}
    \bra{\psi_\gamma} e^{\vec k_1,\sigma_1}_{ij} x^i x^j \ket{\psi_{\beta}} \nonumber\\
    & \times
    \sqrt{\omega_1 \omega_2} 
    \sum_{\tilde\omega_1 = \pm \omega_1}
    \sum_{\tilde\omega_2 = \pm \omega_2}
    \frac{\tilde\omega_1 \tilde\omega_2}{\omega_{\gamma\beta}+\tilde\omega_1} \left(
    \frac{\rme^{\rmi (\omega_{\alpha\gamma} + \tilde\omega_2) t}-1}{\omega_{\alpha\gamma}+\tilde\omega_2}
    - \frac{\rme^{\rmi (\omega_{\alpha\beta} + \tilde\omega_1 + \tilde\omega_2) t}-1}{\omega_{\alpha\beta}+\tilde\omega_1+\tilde\omega_2}
    \right) \nonumber
     \\
    =&  \frac{2 \mu^2 \pi G}{  \hbar V M^2 c^2} 
    \sum_{\gamma} 
    \bra{\psi_\alpha} e^{\vec k_2,\sigma_2}_{kl} x^k x^l \ket{\psi_{\gamma}}
    \bra{\psi_\gamma} e^{\vec k_1,\sigma_1}_{ij} x^i x^j \ket{\psi_{\beta}} \nonumber\\
    & \times
    \sqrt{\omega_1^3 \omega_2^3} 
     \left(
    \frac{\rme^{\rmi (\omega_{\alpha\gamma} - \omega_2) t}-1}{(\omega_{\gamma\beta}- \omega_1)(\omega_{\alpha\gamma}-\omega_2)}
    - \frac{\rme^{\rmi (\omega_{\alpha\beta} -(\omega_1 + \omega_2)) t}-1}{(\omega_{\gamma\beta}- \omega_1)(\omega_{\alpha\beta} -(\omega_1 + \omega_2))} \right) \label{amplitude-squared}
\end{align}
\end{widetext}
where in the last step we are taking into account that only driven frequencies ($\omega_1,\omega_2)$ close to the transition frequencies ($\omega_{\alpha\gamma},\omega_{\gamma \beta}$) will have a probability of causing a transition. I.e., 
\begin{align}
&\omega_1+\omega_{\gamma \beta}>>|\omega_1-\omega_{\gamma \beta}|, \quad\omega_2+\omega_{\alpha \gamma}>>|\omega_2-\omega_{\alpha \gamma}|,\nonumber\\
&\text{and} \;(\omega_1+\omega_2) +\omega_{\alpha \beta}>>|(\omega_1+\omega_2) -\omega_{\alpha \beta}|.\nonumber
\end{align}
The transition probability is obtained as the square magnitude of the above amplitude:
\begin{widetext}
\begin{align}
P_{\alpha\beta}(t) =& 
\abs{a^{(2)}_{\alpha\beta}(t)}^2
\nnl =& \left(\frac{2 \mu^2 \pi G}{  \hbar V M^2 c^2}\right)^2 \omega_1^3 \omega_2^3\;|V_{\alpha \gamma} V_{\gamma \beta}|^2 \;\Bigg( \frac{e^{i\omega_{\delta \gamma}t} - e^{i(\omega_{\alpha \gamma}-\omega_2)t} - e^{-i(\omega_{\alpha \delta}-\omega_2)t} +1}{(\omega_{\gamma \beta}-\omega_1)^2 (\omega_{\alpha \gamma}-\omega_2)^2} \nonumber\\
&- \frac{e^{i(\omega_1 -\omega_{\gamma \beta})t} - e^{i(\omega_{\alpha \gamma}-\omega_2)t} - e^{-i(\omega_{\alpha \beta}-(\omega_1 +\omega_2))t} +1}{(\omega_{\gamma \beta}-\omega_1)^2 (\omega_{\alpha \gamma}-\omega_2)^2} 
- \frac{e^{i(\omega_1 -\omega_{\delta \beta})t} - e^{i(\omega_{\alpha \delta}-\omega_2)t} - e^{-i(\omega_{\alpha \beta}-(\omega_1 +\omega_2))t} +1}{(\omega_{\gamma \beta}-\omega_1)^2 (\omega_{\alpha \gamma}-\omega_2)^2}\nonumber \\
&+\frac{\sin^2{[(\omega_{\alpha \beta}-(\omega_2+\omega_1))t/2]}}{(\omega_{\gamma \beta}-\omega_1)(\omega_{\delta \beta}-\omega_1) (\omega_{\alpha \beta}-(\omega_2+\omega_1))^2}\Bigg).
\label{probability}
\end{align}
\end{widetext} with $|V_{\alpha \gamma} V_{\gamma \beta}|^2 =V_{\alpha \gamma} V_{\gamma \beta} V^*_{\alpha \delta} V^*_{\delta \beta}$, and $V_{\alpha \gamma} = \bra{\psi_{\alpha}} \hat{H}_{int}\ket{\psi_{\gamma}}$.
The transition rate is found by
\begin{align}
    \Gamma = \lim_{t \rightarrow \infty} \frac{P_{\alpha \beta}(t)}{t}.
\end{align}
The time independent terms in $P_{\alpha \beta}(t)$, divided by $t$, as well as the $~  e^{-i\omega t}$ terms will go to zero. The $~e^{+i\omega t}$ terms will cancel each other in the limit. Hence, the remaining term is the last part of (\ref{probability}). By using
\begin{align}
    \lim_{t \rightarrow \infty} \frac{\sin^2{[(\omega_{\alpha \beta}-(\omega_2+\omega_1))t/2]}}{(\omega_{\alpha \beta}-(\omega_2+\omega_1))^2} = \frac{\pi t}{2} \delta(\omega_{\alpha \beta}-(\omega_2+\omega_1)),
\end{align}
we find a constant transition rate
\begin{widetext}
\begin{align}
    \Gamma =& \left(\frac{2 \mu^2 \pi G}{  \hbar V M^2c^2}\right)^2 \omega_1^3 \omega_2^3\;\frac{V_{\alpha \gamma} V_{\gamma \beta} V^*_{\alpha \delta} V^*_{\delta \beta}}{(\omega_{\gamma \beta}-\omega_1)(\omega_{\delta\beta}-\omega_1) }\frac{\pi }{2} \delta(\omega_{\alpha \beta}-(\omega_2+\omega_1)\\
    =& \frac{\pi}{2}\left(\frac{\mu^2 \pi G}{  \hbar V M^2 c^2}\right)^2 \Big|\sqrt{\omega_{\beta \gamma}^3\omega_{\gamma \alpha}^3} \Big|^2  \Bigg|\frac{V_{\alpha \gamma} V_{\gamma \beta}}{\omega_{\gamma \beta}}\Bigg|^2.
\end{align}
\end{widetext}
The delta functions makes it clear there exist one exact final state with energy $\omega_{\alpha \beta}$ and the transition rate is zero for the entire gravitational wave spectrum except for the frequency $\omega \equiv \omega_1+\omega_2 = \omega_{\alpha \beta}$. We can also consider a distribution of final energy states by summing over all possible final states $\alpha$
\begin{widetext}
\begin{align}
    \Gamma =& \frac{\pi }{2} \left(\frac{2 \mu^2 \pi G}{  \hbar V M^2 c^2}\right)^2 \omega_1^3 \omega_2^3\; \sum_{\alpha} \left|\frac{V_{\alpha \gamma} V_{\gamma \beta}}{(\omega_{\gamma \beta}-\omega_1)}\right|^2 \delta(\omega_{\alpha} -\omega_{\beta}-\omega)\\
    =&  2\pi \left(\frac{ \mu^2 \pi G}{  \hbar V M^2 c^2}\right)^2 \omega_1^3 \omega_2^3\; \int d\Tilde{\omega} \; \delta(\Tilde{\omega} - \omega_{\beta} -\omega) \sum_{\alpha} \left| \frac{V_{\alpha \gamma} V_{\gamma \beta}}{(\omega_{\gamma \beta}-\omega_1) }\right|^2 \delta(\Tilde{\omega} -\omega_{\alpha})\\
    =&  2\pi \left(\frac{ \mu^2 \pi G}{  \hbar V M^2 c^2}\right)^2 \omega_1^3 \omega_2^3\;  \left| \frac{V_{\alpha \gamma} V_{\gamma \beta}}{(\omega_{\gamma \beta}-\omega_1) }\right|^2 \int d\Tilde{\omega} \; \delta(\Tilde{\omega} - \omega_{\beta} -\omega)  \rho(\Tilde{\omega})\\
    =&  2\pi \left(\frac{\mu^2 \pi G}{  \hbar V M^2 c^2}\right)^2 \omega_1^3 \omega_2^3\;  \left| \frac{V_{\alpha \gamma} V_{\gamma \beta}}{(\omega_{\gamma \beta}-\omega_1) }\right|^2 \rho(\omega_{\alpha}),
\end{align}
\end{widetext}
where we can recognize the density of states$$ \rho(E) =\frac{1}{\hbar} \rho(\Tilde{\omega}) =\frac{1}{\hbar} \sum_{\alpha} \delta(\Tilde{\omega} - \omega_{\alpha}),$$ considered at the final frequency $\omega_{\alpha} = \omega_{\beta} + \omega $, and we assumed that $V_{\alpha \gamma}$ is approximately constant over the range of the final state distribution.

Next, we wish to solve $V_{\alpha \gamma}$ and   $V_{\gamma \beta}$ similar to the electrodynamic case above. The quadrupole moment is written in a basis of rank two traceless tensors related to the $l=2$ spherical harmonics. I.e., the spherical harmonics related to the spin$-2$ field. Recall the $l=2$ spherical harmonics,
\begin{align}
    Y^{22}&=\sqrt{\frac{15}{32 \pi}} (e^{i\phi} \sin{\theta})^2\\
    Y^{21}&=-\sqrt{\frac{15}{8 \pi}} e^{i\phi} \sin{\theta} \cos{\theta}\\
    Y^{20}&=\sqrt{\frac{5}{16 \pi}} (3 \cos{\theta}^2-1),
\end{align}
and the unit vector $\hat{r}$ is written in polar coordinates as $r_x = \sin{\theta} \cos{\phi}, r_y = \sin{\theta} \sin{\phi}, $ and $r_z = \cos{\theta}$. Plugging the unit vector in the expressions for $Y^{2m},$ and we find 
\begin{align}
Y^{2m} = \mathcal{Y}^{m}_{ij} r^i r^j \rightarrow r^i r^j -\frac{1}{3}\delta_{ij}= \sum_{m} \mathcal{Y}^{m *}_{ij} Y^{2m}.\label{eqn:basis spherical harmonic}
\end{align}
Here, $\mathcal{Y}^{m}_{ij}$ are the spherical unit tensors similar to the spherical unit vectors $\hat{u}_m$ in the electrodynamics case and can be found by solving equation (\ref{eqn:basis spherical harmonic}). This results in
\begin{align}
    \mathcal{Y}^{\pm 2}_{ij} &=
    \begin{pmatrix}
        1 & \pm i & 0\\
        \pm i & -1 & 0\\
        0 &0 & 0
    \end{pmatrix},\\
     \mathcal{Y}^{\pm 1}_{ij} &=
    \begin{pmatrix}
        0 & 0 & 1\\
        0 & 0 & \pm i\\
        1 & \pm i & 0
    \end{pmatrix},\\
      \mathcal{Y}^{0}_{ij} &=
    \begin{pmatrix}
        -1 & 0 & 0\\
        0 & -1 & 0\\
        0 &0 & 2
    \end{pmatrix}.
\end{align}
The system in spherical coordinates is given by
\begin{align}
    r_ir_j -\frac{1}{3}\delta_{ij}=& \frac{8 \pi}{15} [ \mathcal{Y}^{2*}_{ij} Y^2_2 + \mathcal{Y}^{-2*}_{ij} Y^{-2}_2 + \mathcal{Y}^{0*}_{ij} Y_2^0 \nonumber\\
    &+ \mathcal{Y}^{1*}_{ij} Y_2^1+\mathcal{Y}^{-1*}_{ij} Y_2^{-1}].
\end{align}
By computing the transition matrix, we can see that the mass quadrupole interaction is associated with $l=2$ transitions and zero for a $l=1$ transition. Without loss of generality, we consider a gravitational wave traveling in $\hat{z}-$ direction. For the $+$ polarization, we have
\begin{align}
 & \bra{\psi_\alpha}  r^ir^j e_{ij}^{k_z,+}\ket{\psi_{\beta}} = \bra{1s} |r|^2 \ket{2 p_0}\bra{s_0} (r_{xx}^2-r_{yy}^2) \ket{p_0}\\
  &= \frac{8 \pi}{15}\bra{1s} |r|^2 \ket{2 p_0} \bra{s_0}(Y_2^2 + Y_2^{-2})\ket{p_0}.
\end{align}
The angular part is
\begin{align}
    &\frac{8 \pi}{15}\bra{s_0}(Y_2^2 + Y_2^{-2})\ket{p_0}\\
    &= \frac{\sqrt{3}}{4\pi} \frac{16 \pi}{15}\sqrt{\frac{15}{32 \pi}} \int_0^{2\pi} \int_0^{\pi}\sin^4{\theta} \cos{\theta} \cos{2\phi} \;d\theta d\phi,
\end{align}
where the $\phi$ integral is zero. In other words, a direct transition is not possible.
To find all the possible transitions via an intermediate eigenstate, we can evaluate the $\phi$ integral. Note that
\begin{equation}
    \int_0^{2\pi} e^{i n \phi} \cos{2\phi} \; d\phi =  
    \begin{cases}
      \pi & \text{if $n = \pm 2$}\\
      0 & \text{if $n \neq \pm 2$}
    \end{cases}.
    \label{integral phi}
\end{equation}
Hence the eigenstates of the Hydrogen-like atom, which have a wavefunction $\psi_{\gamma} \propto e^{i n\phi}$, will add an intermediate part $\ket{\psi_{\gamma}} \bra{\psi_{\gamma}} \propto e^{i n\phi} e^{-i n\phi} =1$.

For the $\times$ polarization, we have
\begin{align}
 & \bra{\psi_\alpha}  r^ir^j e_{ij}^{k_z,\times}\ket{\psi_{\beta}} = \bra{1s} |r|^2 \ket{2 p_0}\bra{s_0} 2 r_{xy} \ket{p_0}\\
  &= \frac{- 16 i \pi}{15}\bra{1s} |r|^2 \ket{2 p_0} \bra{s_0}(Y_2^2 - Y_2^{-2})\ket{p_0}.
\end{align}
The angular part is
\begin{align}
    &\frac{- 16 i \pi}{15}\bra{s_0}(Y_2^2 - Y_2^{-2})\ket{p_0}\\
    &= \frac{\sqrt{3}}{4\pi} \frac{32 \pi}{15}\sqrt{\frac{15}{32 \pi}} \int_0^{2\pi} \int_0^{\pi}\sin^4{\theta} \cos{\theta} \sin{2\phi} \;d\theta d\phi,
\end{align}
where the $\phi$ integral is once again zero. Similar to (\ref{integral phi}) we have
\begin{equation}
    \int_0^{2\pi} e^{i n \phi} \sin{2\phi} \; d\phi =  
    \begin{cases}
      i \pi & \text{if $n = \pm 2$}\\
      0 & \text{if $n \neq \pm 2$}
    \end{cases}.
\end{equation}

\bibliography{localbib}

\end{document}